# Prolegomena To Any Future Device Physics

Adam L. Friedman and Aubrey T. Hanbicki

***Abstract*—For the last 60 years, advances of computing platforms have been driven by Moore's law. In its essence, Moore's law is a description of the exponential increase in transistor density acting as a proxy for computing power as function of time. While Moore's law began as an interesting observation, it has evolved into a self-fulfilling prophecy used to drive the entire semiconductor industry. Arguments for or against the end of Moore's law have proliferated, and the reluctant consensus is that Moore's law will disappear. Its end is repeatedly thwarted by advances in many different aspects of the computing ecosystem including materials improvements, device design, device/circuit "cleverness," and software and architectural innovations. While many argue that the impending doom of Moore's law is the ultimate roadblock imposed by atomic length scales, quantum processes, and energy consumption limits, we contend that Moore's law must be jettisoned for a different reason: Words matter. Even those who adamantly declare the end of Moore's law still use the language of Moore's law. The inward focus imposes an intellectual tyranny that inhibits revolutionary progress. We instead suggest a more outwardly focused perspective and a shift in language to a regime we coin the Feynman Mandate. In this perspective, we outline the issues with the continued use of Moore's law as well as a prescription for transitioning to a new lexicon. Rather than imposing a new framework wholesale, our intention is to start a discussion.***

***Index Terms*—Beyond CMOS, Computing, Devices, Moore's Law, next-generation computing, Novel materials and devices, Philosophy of Science**

A.L. Friedman is with the Laboratory for Physical Sciences, 8050 Greenmead Dr., College Park, MD 20740 USA (email: afriedman@lps.umd.edu).

A.T. Hanbicki is with the Laboratory for Physical Sciences, 8050 Greenmead Dr., College Park, MD 20740 USA (email: hanbicki@lps.umd.edu).

## I. Introduction

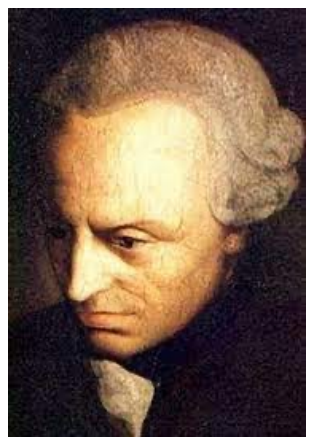

*Human reason is so keen on building that more than once it has erected a tower, and afterwards has torn it down again in order to see how well constituted its foundation may have been. It is never too late to become reasonable and wise; but if the insight comes late, it is always harder to bring it into play.*
**Immanuel Kant, "Prolegomena to Any Future Metaphysics"**

In his *Prolegomena to Any Future Metaphysics*,[1] philosopher Immanuel Kant laid out conditions necessary to enable future philosophical discussion. He realized that the previous bottom-up approaches to "science" must also contend with top-down approaches. That is, learning what exists from basic principles must be congruent with objective reality, and basic principles must likewise be derivable from what exists. A condition essential to merging approaches such as these, and achieving any new paradigm, is a shift in language. Words and definitions are critical because they reflect overall thinking and understanding, and the proper lexicon can drive innovation by appropriately describing problems and framing solutions.

Although we are certainly not the first to tread the "beyond Moore's law" pathway, we want to begin a new type of discussion on the future of computing based on logical philosophy. Kant states, "…things cannot possibly remain on their present footing. It seems almost laughable that, while every other science makes continuous progress, metaphysics…perpetually turns round on the same spot without coming a step further."[2] Of course while Kant was concerned with knowledge itself, we are concerned with an admittedly more modest subject, progress in computing. Humility aside, the future of computing continues to have significant societal impact, and, while progress has seemingly continued apace, we contend that, as long as progress remains tethered to the language of Moore's law, soon we will likewise be doomed to turn round without coming a step further. Although many might say there is agreement that Moore's law is dead, a survey of the literature corroborates that this is a reluctant consensus at best. If Moore's law is truly dead, why do essays about it continue to proliferate [3]?

To be fair, many positive steps toward jettisoning Moore's law abound. For instance, the notion of the end of computing nodes in favor of more holistic integration to end bottlenecks, termed "hyper-scaling," is gaining traction [4]. Nonetheless, many computing consortiums continue to argue amongst themselves without firmly adopting the language and intent of these new paradigms, for example, the Rebooting Computing Initiative.[5] We contend that brain energy would be better utilized if all discussion of Moore's law is abandoned in favor of more holistic, outward looking language that better encompasses the future.

Moore's law began as an observation of the density of transistors in a circuit. It was subsequently transformed into a self-fulfilling prophecy, serving as a roadmap for the entire semiconductor industry.[6] It has been re-imagined, re-invented, and ultimately embraced as a mindset, an approach, and a philosophy. While it certainly has been a simple, useful guidepost, it is an inward-focused model with a clear finish line, and its time has passed. We have accomplished everything envisioned by Gordon Moore in 1965 [7] and are at the logical end to this path.

Clarion calls announcing the end of Moore's law have been a staple of the industry for over 20 years (Fig. 1). Through the years, hurdles were cleared and roadblocks circumvented through a variety of ingenious strategies, what Moore termed "device and circuit cleverness."[8] Nevertheless, the hard reality of the atomic limit still looms. While significant effort is made on "more Moore" despite the impending terminus, there is also progress toward a realm dubbed "beyond Moore." However, we assert that both are, in actuality, "Moore" of the same. By framing the problem in terms of Moore's law, there is no shift to another paradigm. We are stuck and remain without any central guiding principle to outline what comes next. No Moore!

Some have posited corollaries and new, named trend "laws" such as Dennard scaling [9] and Koomey's law [10]. The similarity in construction of these scaling relations to Moore's law (all are generalized exponentials) inhibits the adoption of any new jargon because they do not actually create new intellectual paths. The authors of Koomey's law, as well as others [11], seem to understand this deficiency. Ultimately, scientific paradigms are defined by agreed upon rules, definitions, and standards for scientists in a particular field.[12] A paradigm can transform only through a revolution and it is past time to revolt.[13]

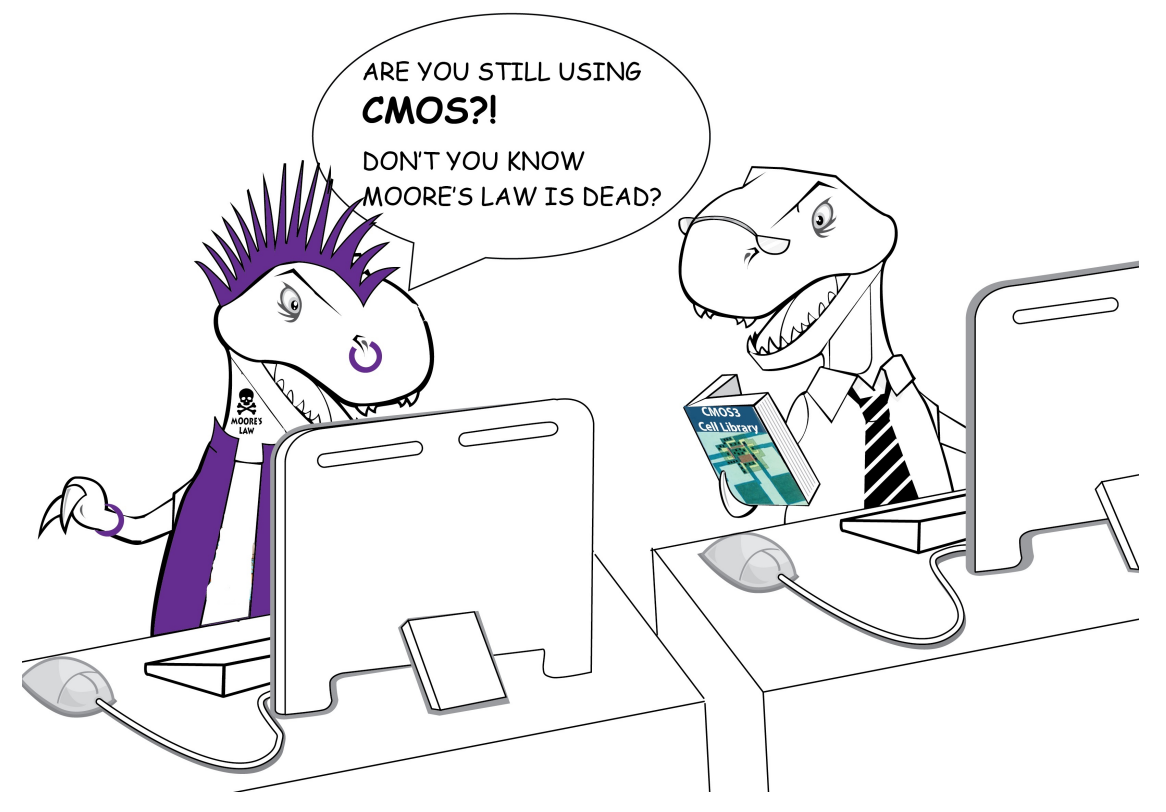

Figure 1: The end of Moore's law has been a source of many memes.

One possible path toward a new scientific paradigm originates to before the 1965 observation by Moore. Richard Feynman, in his famous speech, "There's plenty of room at the bottom,"[14] given at the American Physical Society meeting in December 1959, is often credited with inventing the field of nanoscience. He revisited a similar line of reasoning in a 1983 speech given at the Jet Propulsion Laboratory and, astonishingly, established the field of quantum computing.[15] Although his prophesies do not specify a pathway to advance classical computing, Feynman does establish a way forward by suggesting a holistic, outward-looking approach to device design. This approach is nicely summarized by his comment, "It would be interesting in surgery if you could swallow the surgeon." In essence, *we should consider the problem in its entirety to creatively formulate unique solutions* rather than rely on dogmatism. We brand this approach as "Feynman's Mandate."

In this perspective, we outline the intellectual deficiencies with Moore's law and why it should finally be surrendered as a roadmap for future computing. We introduce and discuss the idea of a "Feynman's Mandate" era serving as a guidepost for the next paradigm. This new age is heavily reliant on holistic co-design encompassing the entire computing ecosystem from materials to devices to architectures to software. We give examples of how and where we believe this paradigm has already succeeded. Mostly we want to begin a discussion that that will enable a new outlook to evolve and be adapted synergistically. It is not our intention to present and impose a fully formed set of principles, metrics and language, but rather to start a conversation in the community.

## II. THE PROBLEM WITH MORE MOORE

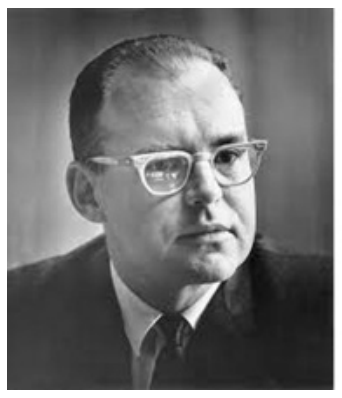

*We seek durable change, not simply delaying consequences for a short time.*
**Gordon and Betty Moore Foundation**

Let us reiterate upfront that ours is not a standard critique of Moore's law. We understand that ingenuity with materials and integration, for example incremental engineering optimization with device designs (*e.g.,* MOSFET to finFET) and/or architecture (*e.g.*, increased computational cores and more efficient parallel processing), can sustain parts of Moore's law for a while longer. We contend that the most damaging aspect of the continued adherence to Moore's law is intellectual oppression. Let us begin by understanding what Moore's law *is*: In its essence, Moore's law is an empirical observation of a logarithmic trend. The density of devices in a circuit doubles roughly every two years. The originator of this "law," Gordon Moore, a co-founder of Intel Corporation, actually had an initial formulation that he then modified after further empirical input. In 1965, the observation and prediction (Fig. 2a) was an annual doubling of transistor densities.[7] In 1975,[8] having been satisfied with 10 years of data indicating exponential growth, he revised the doubling period to two years. Subsequent formulations, corollaries, and scaling laws abound.

In 1965, Moore noted that the trend in device density was driven by the microelectronics community whose primary goal at that time was to add more components to circuits while keeping the cost per component at a minimum (Fig. 2b). The competing effects of increased component density and decreased yield defined the trajectory. In 1975, a decade after his first insights, he highlights "complexity" of integrated circuits, which is more inclusive of the different incorporated devices (Fig. 2c). His new analysis, Fig. 2d, included the familiar device scaling trends, a concept we closely associate with Moore, but also a more nebulous "device and circuit cleverness" factor. While device scaling is synonymous with Moore (Fig 2d), "device and circuit cleverness" has been paramount to enabling obedience to Moore's law.

The long-lasting adherence to Moore's law has habituated its use as a benchmark. Indeed, its simplicity likely explains its 60-year(!) endurance; the perception that easier to understand ideas are more highly valued is a well-known phenomenon in psychology called the fluency heuristic.[16] Nonetheless, maintaining a simple metric simply because it is simple is illogical. Moore's law is no longer an interesting observation, it is a sacred standard. Aptly named E.P. DeBenedictis, writing for the Rebooting Computing Initiative[5] describes an almost biblical stance: "Moore's law is a fluid idea whose definition changes over time. It thus doesn't have the ability to 'end.'"[17] Aside from being incorrect (by definition, a law is tied to a particular observation and if that observation does not hold up, the law does not either), we hypothesize that this evolving definition is due to the fluency heuristic. People are loath to throw it out even when it makes sense to do so.

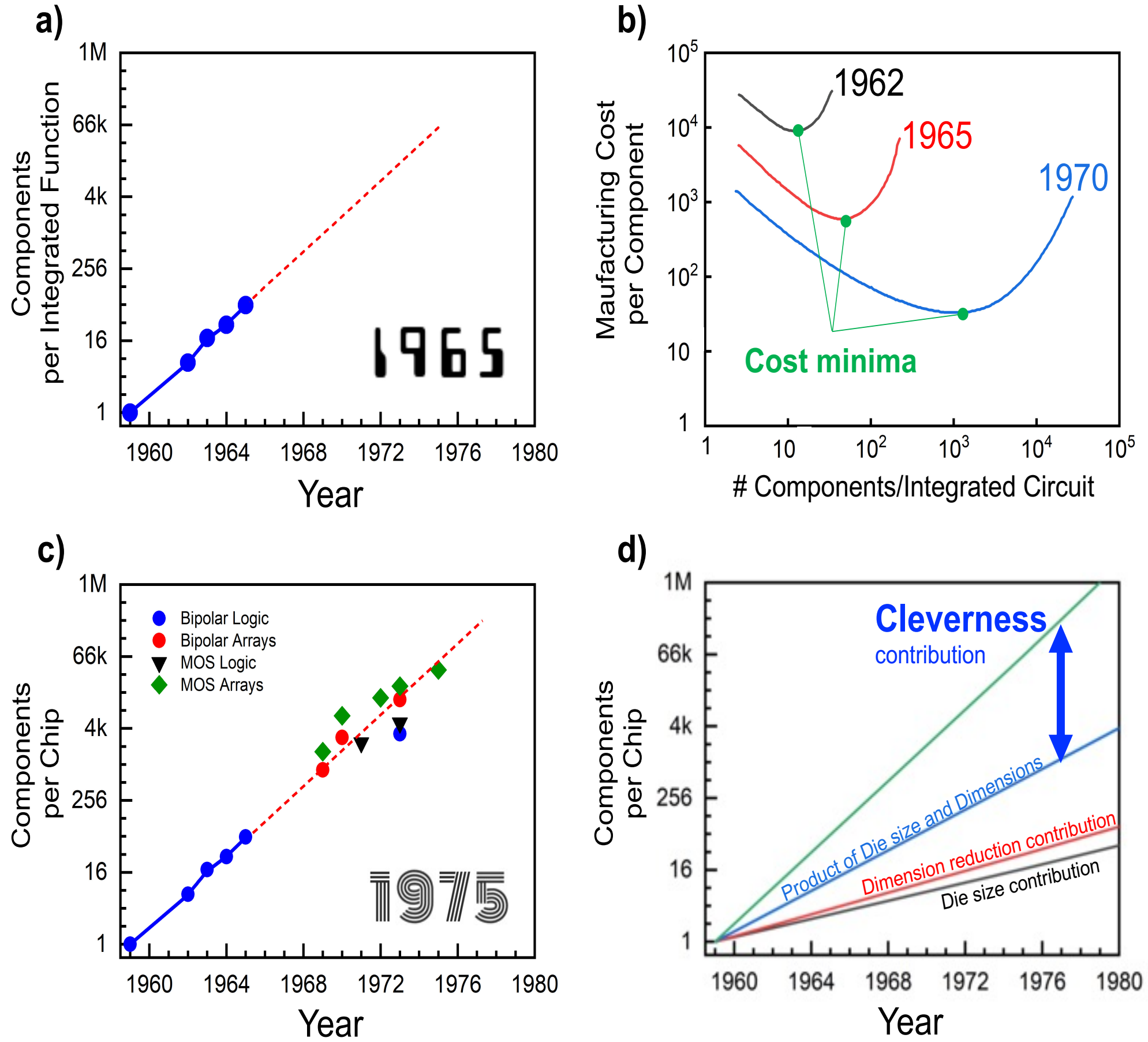

Figure 2: The origins of Moore's law. Presented in 1965, a) and b) are the original construction of Moore's law. They consist of a) the original logarithmic dependence of component count on year and b) cost per component minima curves used to predict the logarithmic trend. A new analysis in 1975 included c) an updated logarithmic trend included various types of novel computing elements and d) a breakdown of the factors contributing to the exponential dependence. Note the substantial contribution of the nebulous quantity called "cleverness." All data extracted from references [7] and [8].

Moore's law remains in its most basic definition an observation/goal of computing growth, and calling it a "law" is specious. It would better be described as a regularity, and historically, exponential growth curves are frequently used to demonstrate regularities.[18] Indeed, given a sufficient time scale, every technology appears to grow exponentially.[19] Such curves are easy to understand and convey a fantastic sense of change. Hence, they are particularly susceptible to the fluency heuristic.

Although the analysis of exponential trends is useful to organize data and help visualize overarching trends, can a real understanding of the future of computing be extrapolated and sustained from a simple exponential tendency? In fact, technology growth has been super-exponential rather than exponential. Thus, even the premise is flawed.[20] Moreover, computer performance is really measured using elaborate benchmarks such as those developed by SPEC.[21] In practice, how efficiently problems are solved, and not the number of transistors, is what matters. We enumerate additional troubling deficiencies with Moore's law below.

(1) Metrics of Technology Must Adapt: Computing technology has evolved from the abacuses, to counting boards, to difference engines, to calculators, to punch card readers, *etc.* Nowadays, the commonly used metric to measure computing power is computations per second (CPS) or millions of instructions per second (MIPS).[22] This metric works well when the goal is, roughly, arithmetic. However, future goals will not be based on MIPS. One can see this in complicated computing formats such as those based on cloud or edge computing [23]. Future metrics will be based on the efficiency of tackling specific problems. Consider for instance solving the traveling salesman problem with a conventional, von Neumann-based system *vs*. a quantum computer.[24] Transistor density is no longer an approximation of computing competency.

(2) Intellectual Tyranny: We use this hyperbole to highlight how trying to align and define all progress using Moore's law stifles creative solutions and out-of-the-box thinking. It may be satisfying to map progress on a single curve, but we are scientists and not poets and need not be limited by perceived aesthetic value.[25]

Furthermore, this objection falls into the realm of what the psychology of science knows as "preregistration."[26] Preregistration involves the detailing and vetting of hypotheses and plans before performing experiments. When combined with a rigorous verification cycle, preregistration should aid in the creativity process. However, it breaks down when the original analysis plan needs to be adjusted. If Moore's law is considered the hypothesis, then the plan is contained in the various forward-thinking roadmaps used by those in the field such as the IDRS[27] or the Decadal Plan.[28] But, Moore's law is not a hypothesis or a law in any sense of the word. It is an observation made *a posteriori* and extrapolated, assuming that the conditions that precipitated it continue to be true. Rather than engage in a cycle of rigorous verification, we blindly accept a faulty hypothesis, which runs counter to our training as scientists.

(3) Expectations Impact Reality: When we force the model to fit reality, our perceived reality can be biased. This is an example of extreme confirmation bias [29]. As stated above, Moore's law is not a hypothesis. However, we can derive a hypothesis from its sentiments: "The number of transistors on a chip will continue to double approximately every two years for the foreseeable future and this is a good measure of computer performance." We can address both parts of this separately. (1) The number of transistors on a chip cannot continue to double in the same way (the end of the node [4]). So, this hypothesis cannot possibly hold. (2) the number of transistors on a chip is only a good measure of performance for von Neumann and von Neumann-like architectures (for example multi-core, parallel computing). So, this part of the hypothesis also fails both in conjunction with the first part and by itself. Trying to confirm this faulty hypothesis, we continue along the well-trodden path of confirmation bias.

The so-called Principal of Verification [25] posits science can only debate statements that can be proven true or false using empirical logic and scientific methodology. Because Moore's law can warp itself so that someone could always find a way that it looks like its principles are satisfied, there would be no way to fully determine that it was actually satisfied. If the hypothesis cannot be fully tested, it fails to be a hypothesis, and is non-sensical—a logical fallacy.

(4) Physical Limitations Must Be Reckoned With: By refusing to think creatively or plan, we simply kick the can down the road. Indeed, not addressing the long term "what's next" questions could become catastrophic. In reference to his work with the Gordon and Betty Moore Foundation, Moore himself stated that "we seek durable change, not simply delaying consequences for a short time."[30] One impending physical limit, energy consumption, is highlighted in an SIA/SRC report[31] and echoed by the Decadal Plan [28]. While savings can be found across the board (barrier height, devices, I/O, and circuits), Moore's law does not acknowledge energy consumption as a cost. Naysayers may reference Koomey's law again [10], but this is just Moore restated. Energy is not an explicit problem here, it is just a way to track computational progress, which of course grows naturally over time, without prescribing paths toward new architectures or cleverness that could subvert crisis. Furthermore, some in the high-performance computing community agree that such "laws" do not adequately describe the necessary balance between performance and intermural, across-the-board energy consumption [11]. A Department of Energy projection shows that the world's computing systems will have power needs that outstrip supply by approximately 2037 on the current trajectory. Existing "More Moore" solutions will only delay this date by approximately 5 years. [28,31] These projections do not even include the explosion of demand generated by artificial intelligence [32]. While the demand is being met with novel solutions such as FPGAs and GPUs, it is a conventional response and does nothing to mitigate the demand in energy. One estimate based on a forecast by the International Energy Agency (IEA) is the AI industry is expected to increase its energy demand by an order of magnitude from 2023 to 2026. [33] All of these energy projections also do not include any other energy consumption besides that required for computation, e.g. cooling, maintenance and up-keep, access, and fabrication.

(5) Enforced Silo-ing: Moore's law does not address the computational system in its entirety, and does not consider other systems that could be better at solving certain problems.[34] This consideration is exceedingly important because performance is not limited by number of transistors on a chip but various other systemwide bottlenecks. As one example, interconnects are a well-known bottleneck left unaddressed by Moore's law that are more problematic with increased size scaling.[35, 36] How will this affect future computing systems such as neuromorphic architectures where interconnects between device neurons are essential? There are ideas for addressing it using integrated photonics,[37] which has no place in Moore's law at all! We must also contend with the bandwidth bottleneck,[38] von Neumann bottleneck,[39] energy bottleneck, software efficiency, *etc*.. Many of these performance-limiting bottlenecks have been known for a very long time and have nothing to do with transistor density. Indeed, CPU clock speeds leveled off in about 2005![28]

We need to understand exactly the types of problems that need to be addressed and design a system that efficiently addresses them. This current and future growth in complexity of computing systems is not accurately enough summarized by Moore's law.

## III. LOOKING FORWARD: FEYNMAN'S MANDATE

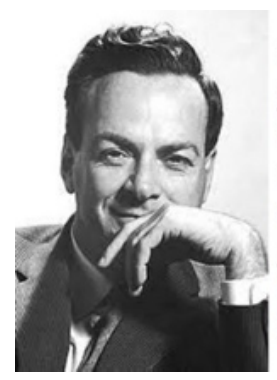

*What could we do with layered structures with just the right layers? What would the properties of materials be if we could really arrange the atoms the way we want them? They would be very interesting to investigate theoretically. I can't see exactly what would happen, but I can hardly doubt that when we have some control of the arrangement of things on a small scale we will get an enormously greater range of possible properties that substances can have, and of different things that we can do.*
**Richard Feynman, "There's plenty of room at the bottom," APS Meeting December 29, 1959.**

Now comes the hard part. We have discussed why using the phrase "Moore's law" is limiting. In fact, all language is limiting at some level. Nonetheless, it is certainly helpful to have a guiding principle for success with the appropriate language to help propel these tenets forward in a broadly applicable way. We contend that any new vocabulary must implicitly and explicitly consider advances in materials, device design, interconnects, architecture, computing platforms, and hopefully even parameters yet unknown. While "Moore's Law" was a useful rubric for many years, its utility was ultimately always going to be limited by its focus on only one specific aspect of the computing ecosystem. Going forward, we require a more encompassing, inevitably more *abstract*, outward looking vision (Fig. 3).

True abstractions are not necessarily conducive to focus, however. Therefore, we would like to introduce a concrete notion we will call "Feynman's Mandate," named for the ideas demonstrated in Richard Feynman's famous address. While the specific approach suggested by Feynman, "atom-by-atom assembly," certainly contributed to the field of nanotechnology, it is more the broad applicability that we want to tap into here. By looking into the future, he establishes a general mindset that *considers the problem in its entirety to creatively formulate a unique solution*. For computing, language adopting Feynman's Mandate assumes broad synergy between all the relevant computing pieces and performers and creativity to solve difficult problems. An approach based on Feynman's mandate is not stuck on a single dying metric and outdated language. Rather, it changes to encompass the vision of the future (not being a law, it is *allowed* to change) and acts as a mindset.

How do we swap the hyper-focused "Moore's law" for the more outward looking "Feynman's Mandate?" Often, this is already the way science works, or at least this is how science *should* work. Researchers from such varied disciplines as Physics, Chemistry, Computer Science, Materials Science, Electrical Engineering, *etc.*, all contribute substantively to collective progress. Certainly, we need to focus on devices, but CMOS and the transistor are not the only games in town. Indeed, research into novel devices structures, those based on alternate state variables like spin, valley, and phase has been ongoing for decades. Co-design of systems, where the properties of devices are tailored to novel computing platforms such as multi-valued logic and neuromorphic computing is likewise gaining traction. In fact, discussions of more Moore always seem to include co-design that relies on systemic integration rather than simply cramming more transistors.[40] But, to let these new technologies fully develop, this holistic, systemic approach needs to be reflected in the language we use.

Steps are already being made in this direction. S. K. Moore (perhaps a distant cousin?), writing for *IEEE Spectrum*, uncovers a cabal of scientists working to discard the false narrative of nodes and to develop a new metric.[41] T.E. Conte, *et al.*, note that considering co-design beyond Moore's law becomes increasingly disruptive by incorporating systemic changes.[42] Here "disruptive" can be taken to mean both a great change to the current paradigm and/or increasingly difficult to accomplish, perhaps due to embedded psychological preferences. *i.e.*, this new road is very difficult because there is so much to know, and it is decidedly easier to stay in a particular silo.

*Words matter*. As a first step we need to abandon language that intentionally keeps us disconnected. Below, we highlight some additional considerations.

(1) A holistic approach to device concepts must be encouraged and maintained: What we can understand and accomplish with materials and devices has greatly increased in complexity. Ironically, the history and evolution of CMOS is a nice example of this paradigm with each component of the transistor changing over the years. The metal gate electrode (M) evolved from Al to polysilicon to refractory metals, the oxide (O) evolved from silicon dioxide to high-k dielectrics based on hafnia, and the semiconductor (S) has cycled through various preparations and alloys of silicon. Cleverness in incorporating new materials into new device designs is a hallmark of the industry and serves more than just Moore.

A culture based on Feynman's Mandate will embrace technologies that do more than simply shrink the node size, for example, memory and logic elements using novel state variables. One such technology, of many, is the magnetic tunnel junction (MTJ). An MTJ is an element that consists of several layers of magnetic material separated by a thin, insulating layer. The relative orientations of the magnetic layers, *e.g.* parallel or antiparallel, dictates the state of the element. Switching the state of the MTJ is accomplished in commercialized MRAM by a current carrying wire near enough to the MTJ to impart a magnetic field or by using spin transfer torque. Newer concepts like spin-orbit torque could provide a path for even faster and more energy efficient switching enabling computing outside of the traditional node-reduction paradigm.[43] And MTJs can even be used beyond the von Neumann infrastructure as they have been suggested to be suitable for neuromorphic computing.[44]

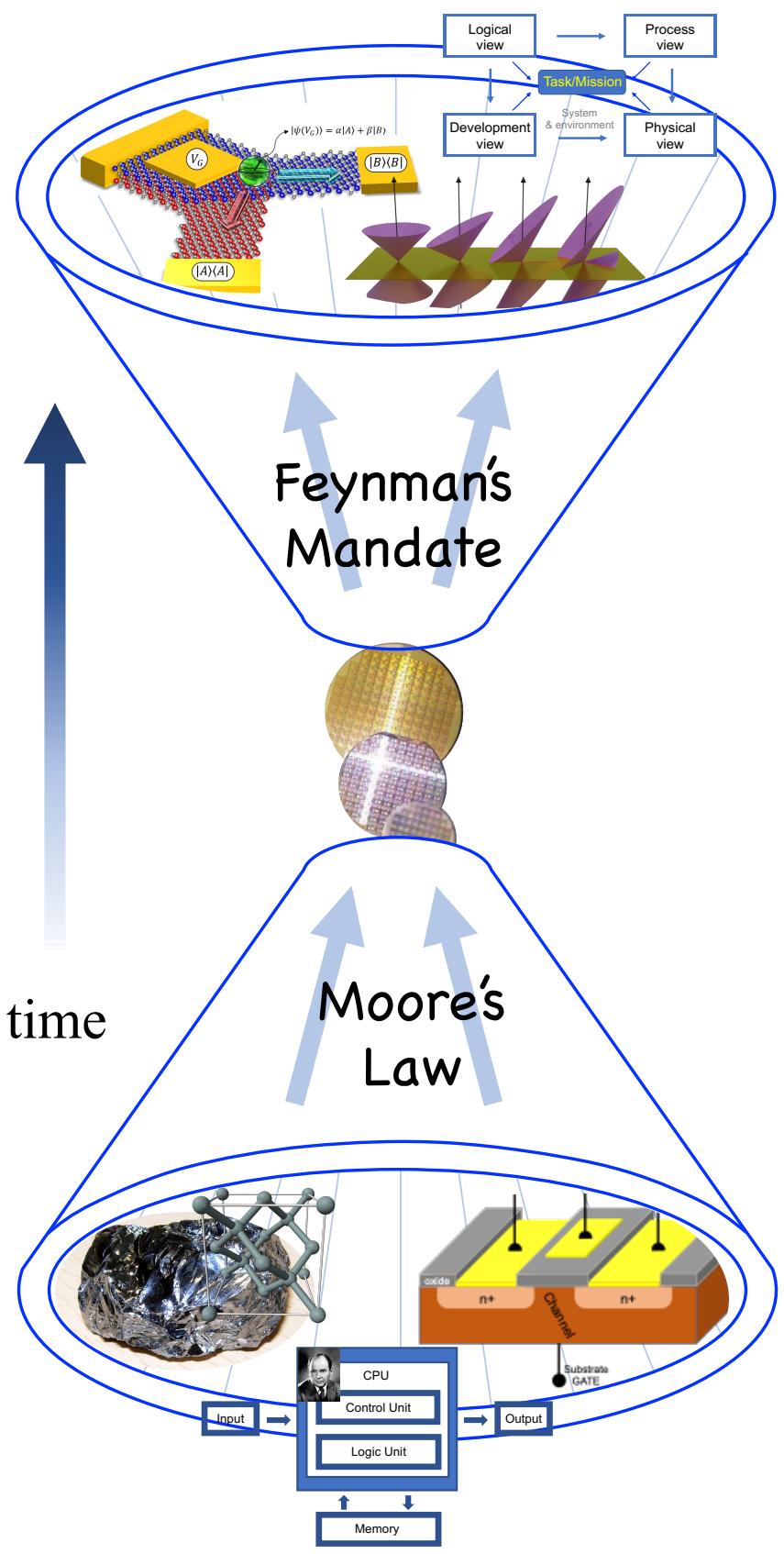

Figure 3: Moore's law has been driven by silicon, the transistor and the von Neumann architecture to focus the semiconductor industry for the last 60 years. This hyper focused approach must now give way to a new paradigm, Feynman's Mandate, where an outward approach is taken to enable, empower and utilize new materials, devices, and architectures.

(2) Humbly preaching to the choir, there needs to be increased synergy between device and architecture design: In addition to considering device design and materials in an integrated way, devices must then also be matched and optimized to particular architectures. One example is using non-volatile memories within a neuromorphic computing platform.[45] As just discussed, MTJs are a specific possibility. Another exciting example is in the push to create large 3D heterogeneously integrated circuits that uses advanced interconnects, packaging solutions, novel nanotechnologies [46], and even additive manufacturing exemplified by the N3XT initiative [47] and the continually discussed in the IEEE Heterogeneous Integration Roadmap [48]. Optimizing components into unconventional circuit geometries, a task that is undoubtedly familiar to practitioners of origami [49,50], while optimizing for efficiency using advanced AI embedded in electronic design automation tools [51] is often described as a 3D extension of Moore's law. Indeed, this is the sort of cleverness that we think Moore would have referred to in Fig. 2. However, we believe that it better fits into our notion of Feynman's mandate where we have cast off the strictures of outdated and thinking for a more creative future. Furthermore, we need to use different architectures for different computing applications. Obviously, von Neumann architecture still works well for many things. Maybe quantum computing or neuromorphic architecture works better for something else. We then would need inter-architectural interconnects, and of course, software design and optimization are also crucial components.

(3) Any new metric of progress for computing must include complexity or "cleverness" while remaining simple enough to be universally understood. Although we may not be the most well-suited authors of a new metric, we suggest that a new metric should include total energy efficiency per mass or volume. It should also include energy cost, both physical and economic. Qualitatively, it could look something like:

$$\text{Computing Prowess} \sim \text{Complexity} / \text{total cost (energy x dollars)}$$

By complexity, we refer to the computational complexity of the problem that one is attempting to solve (so-called "big O notation," or the like). Complexity inherently includes some measure of time to complete the problem at hand, perhaps in computing cycles. In cost, we suggest using a product of the energy expended in building the computer then computing the problem and the dollars expended in building the computer and powering it through the problem. Some of these data are not widely available. Perhaps, for now, we would just use the amount of resources required for operation and in the future expand on data capture. Therefore, this metric gives a complete picture of the entire computing system applied to a particular problem and creates a better way to make comparisons. We suggest naming the unit of (1/J$) a Feynman to further embed Feynman's Mandate in our collective psyche and to acknowledge the scientific prophesies that inspired so many.

For a simplified application of this new metric, we consider the AlphaGo computation *vs.* Go master Lee Sedol. As both AlphaGo and Lee Sedol competed on the same problem, we can normalize the complexity to 1. The Google AlphaGo program on the DeepMind project computer cluster arranged in a neuromorphic architecture used 1,920 CPUs and 280 GPUs filling 20-50 $m^3$.[52,53,54] It also required 1,000 scientists and 80 months (in real time) to train. We make a rough estimate of about $2.2M for the materials cost ($1,000/CPU,GPU) and $600M for people costs (perhaps overestimated, as those 1,000 scientists likely worked on other projects simultaneously, but it serves in this very rough example). With each CPU/GPU using about 400 W and using the 80 months training (the actual game only took a negligible amount of time), we get a power cost of 4.1 $x10^{13}$ J. So the metric would have an upper bound of 1/(602 x $10^6$ * 4.1 x $10^{13}$) = 4.1 x $10^{-23}$ Feynman. AlphaGo defeated Lee Sedol who, according to Google, is 5 ft. 8 inches, was powered by whatever he ate earlier that day, and trained in Go since at least 1997. This would be 216 months of training, but we can assume it was not constant because he needed to sleep and eat. So, we estimate a real time training amount of about 72 months if he trained 8 hours per day every day. A human brain uses about 12 W (Perhaps unfairly, we are ignoring the multitude of years of evolution that produced a brain this efficient). Let us say he costs about $100,000/year for food, etc. We then get a cost of $1.8M and 2.5 $x10^9$ J. The metric would be 2.2 x $10^{-16}$ Feynman. Thus, although Lee Sedol lost

to AlphaGo, the Lee Sedol neuromorphic computer is still significantly more efficient than AlphaGo at this particular problem, occupies significantly less space, and is probably a better choice for most applications. Nonetheless, rather than concentrate on having biological neuromorphic computing 3D printer reactors ("parents") manufacture more Lee Sedol, we can imagine a future AlphaGo that uses significantly less resources such that the metric would be much larger, while the metric for Lee Sedol would not change. We could also then use the metric to compare to different complexity problems, as the result is scaled by the complexity.

(4) Stakeholders have an important role:
Moving into the Feynman Mandate will require the cooperation of all the entities involved in the entire process of computing —academia, government, industry, and end users. Economist Eric von Hippel notes that end users were responsible for most of the major innovations in the semiconductor industry.[55] Nonetheless, major changes to manufacturing processes are hampered by the competing needs of end users and manufacturers, which von Hippel refers to as "stickiness," and others sometimes refer to as levels of disruption in the typical process.[55] This stickiness is likely also one of the culprits contributing to the staying power of Moore's law. Of particular significance to computing, end users want greater efficiency, faster products, more memory, *etc*. But, they do not necessarily care *how* those goals are achieved. So, industry takes the least expensive and perhaps simplest route to the goal, which is to continue making processors and components using the same methods with minimal major changes because this is where they have invested resources and knowledge. This is especially true when a lot of new outside knowledge, such as new materials, state variables, integration, and operating procedures are required. The problem with this sort of stickiness is that, as we have discussed above, they will eventually run out of room and the goals will become impossible to reach. Industry will then be forced to change on a massive scale, which would be prohibitively expensive in a commercial atmosphere dominated by few super-large companies and affect the productivity of all the users down the line.

To understand how the stakeholders should work together on the goal of Feynman's Mandate, we first must take a bird's eye view of the funding landscape. There are effectively four end users: (1) private consumers, (2) government, (3) non-profit (like academia) and (4) commercial. They all usually have very different needs. But, with the exception of private consumers, all are mostly represented by lead-user innovation centers which are large research and development organizations, *e.g.*, federally funded research and development centers (FFRDCs), university affiliated research centers (UARCs), and defense labs for government, university research labs for academia, and internal research and development labs for industry. As far as funders, there are two of consequence: (1) government, and (2) commercial. In the United States, the commercial sector spends significantly more, currently about twice as much, on R&D than government.[56]

The organization of research funds is critical to the advancement of the computing enterprise. The National Science Foundation releases periodic reports that agglomerate and organize the data for U.S. R&D spending. According to those reports, government is more focused on "research" and industry is focused more on "development."[57] Of all basic research, U.S federal spending comprises 41% and industrial spending comprises 30%. For applied research, 34% comes from the government and 54% comes from industry. For experimental development, 12% comes from the government and 86% comes from industry. More than half of private sector money goes specifically to computing related research. About half of government spending is in defense, but most of the overall spending including defense is toward computing related research. A small portion of all R&D spending (17%) is basic research. Academia spends 46% of these funds and industry spends 29%. Applied research accounts for 19% of the overall R&D spending. Of this 57% is spent by industry, 18% by academia, and federal entities account for 17%. From these data, one can roughly visualize the silo-ing effect in research: Academia becomes the basic research "ideas factory," but without efficient implementation. Industry is focused mainly on commercialization. Although it may take some ideas from academia, most of its ideas originate in its own research laboratories, perhaps causing even further silo-ing by self-determining the needs of the end-user without sufficient input from potential stakeholders beyond capitalist economics. Government is focused mainly on the processes involved with government niche-applicable technologies. Each player has a set of skills that could be much better integrated into a holistic computing landscape: academic idea factory, industrial commercialization, and government with facilitation and usage implementation.

Applying the Feynman Mandate to research spending behaviors, we see there must be a mindful understanding of funding co-design projects not just across technology silos but across lead users. If we assume that the funding profile is self-organized such that those who are receiving the funds are best situated to use them in those particular ways, we must make a conscious effort in building inter-silo connections. Researchers working in parts of the technology chain that generally do not talk with each other need to communicate, *i.e.* physicists need to talk with system integration specialists, *etc*. This would primarily be a task for the funders—government and industry—to make sure funds are properly leveraged to make the most progress and motivate true interdisciplinary research. This is particularly important with technologies that are not end-user specific, *i.e.* not in government niche areas or areas where there can be an overlap in consumer and government needs. One example is in next generation magnetic random-access memories (MRAM).[58] Another example is in quantum computing, where co-design may be essential for development and implementation.[59] We need to freely share knowledge across the board to address the co-design portion of Feynman's Mandate. We must further integrate information and knowledge with economic drivers. The rhetorical framework of a "knowledge-based economy" could be useful in that it speaks to gathering data under a single umbrella, bringing visibility to science and creating new investment opportunities.[60]

## IV. Conclusion

Kant concluded his introduction to his *Prolegomena* with an appeal for interested discussion and a thinly veiled attack on his detractors through Virgil's allegorical agricultural poetry.[61] We likewise make an appeal for discussion: we sincerely hope that our perspective serves as an introduction—a prologue, and facilitates a discussion that will lead to advancement in the field through the better presentation and usage of language, metrics, and co-design ingenuity that enables significant future computational advancement. We shall also conclude with Virgil, however, more positively. Even though scientists stereotypically may not gracefully abide change, they will adapt as they understand it as necessary for progress and for driving innovation:

> *contemplator: aquas dulces et*
> *frondea semper*
> *tecta petunt*
> —Virgil, *Georgica*, IV. 61-62
>
> [Translation]
> *Take note: they are continually*
> *searching for sweet waters*
> *And leafy canopies*

## Acknowledgments

The authors gratefully acknowledge J. Marx for creating the cartoon in Figure 1. The authors gratefully acknowledge supportive discussions with T.S. Salter, M. Raugus, D.J. Mountain, P.M. Campbell, and A. Kott. The authors also gratefully acknowledge the 25% of their junior researchers, post-docs, and students who were convinced to actually read this. The authors further gratefully acknowledge N.A. Blumenschein for being the only member of the aforementioned group who provided comments—a true protector of the hive.

## References

[1] I. Kant. *Prolegomena to Any Future Metaphysics*. Cambridge University Press, 1997.
[2] Ibid. page 6.
[3] Maybe this article will be the last!? We can only hope.
[4] S. Salahuddin, K. Ni, and S. Datta, "The era of hyperscaling in electronics," *Nature Electronics,* vol. 1, pp. 442-450, 2018.
[5] https://rebootingcomputing.ieee.org
[6] See, for example, the National Technology Roadmap for Semiconductors (NTRS), first published by the Semiconductor Research Corporation (SIA) in 1992 after the Microtech 2000 Workshop and Report in 1991 when the SIA technology chair was none other than Gordon Moore. It was followed by the International Technology Roadmap for Semiconductors (ITRS) (www.itrs.net), which existed from 1998-2013 before morphing into the International Roadmap for Devices and Systems (IRDS) (www.irds.ieee.org). For additional history, see also W.J. Spencer and T.E. Seidel, "National technology roadmaps: the U.S. semiconductor experience," in *Proceedings of 4th International Conference on Solid-State and IC Technology,* Beijing, China, 1995, pp. 211-220.
[7] A reprinting of Moore's original paper printed in *Electronics*, 114-117 (1965) can be found in G.E. Moore, "Cramming more components onto integrated circuits," *Proceedings of the IEEE*, vol. 86, no. 1, p. 4, 1998.
[8] G.E. Moore, "Progress in digital integrated electronics," *Technical Digest of the International Electron Devices Meeting*, IEEE, 1975, pp. 11-13.
[9] R. H. Dennard, F. H. Gaensslen, H.-N. Yu, V. L. Rideout, E. Bassous, and A. R. LeBlanc, "Design of ion-implanted MOSFET's with very small physical dimensions," *IEEE Journal of Solid-State Circuits*, vol. 9, no. 5, pp. 256–268, Oct. 1974.
[10] J. Koomey, S. Berard, M. Sanchez, H. Wong. "Implications of historical trends in the electrical efficiency of computing," *IEEE Annals of History of Computing*, vol. 33, no. 3, pp. 46-54, 2011.
[11] B. Subramaniam, W. Saunders, T. Scogland, W.-C. Feng. "Trends in Energy-Efficient computing: a perspective from the Green500," *2013 International Green Computing Conference Proceedings*, pp. 1-8, 2013.
[12] T.S. Kuhn, *The Structure of Scientific Revolutions*. University of Chicago Press, 1996.
[13] With all apologies to the late Gordon Moore and utmost respect for his accomplishments. He will, however, be remembered as a part of history rather than the future: http://www.forbes.com/profile/gordon-moore/?sh=527fac8525ac accessed 7/28/2021
[14] R.P. Feynman, "There's plenty of room at the bottom," *Journal of Microelectromechanical Systems*, vol. 1, no. 1, pp. 60-66, 1992.
[15] R.P. Feynman, "Infinitesimal machinery," *Journal of Microelectromechanical Systems*, vol. 2, no. 1, pp. 4-14, 1993.
[16] K.G. Volz, L.J. Schooler, and D.Y. von Cramon, "It just felt right: the neural correlates of the fluency heuristic," *Consciousness and Cognition*, vol. 19, no. 3, pp. 829-837, 2010.
[17] E.P. DeBenedictis, "It's time to redefine Moore's law again," *Computer*, vol. 50, no. 2, pp. 72-75, 2017.
[18] A. Kott, "Toward universal laws of technology evolution: modeling multi-century advances in mobile direct-fire systems," *Journal of Defense Modeling and Simulation: Applications, Methodology, Technology*, vol. 17, no. 4, pp. 373-388, 2020.
[19] A. Kott, P. Perconti, and N. Leslie, "Discovering a regularity: the case of an 800-year law of advances in small-arms technologies," arXiv: 1908.03435, 2019.
[20] B. Nagy, J.D. Farmer, J.E. Trancik, J.P Gonzales, "Superexponential long-term trends in information technology," *Technological forecasting and social change*, vol. 78, no. 8, pp. 1356-1364, 2011.
[21] http://www.spec.org
[22] W.D. Nordhaus, "Two Centuries of productivity growth in computing," *Journal of Economic History*, vol. 67, no. 1, pp. 128-159, 2007.

[23] M.S. Aslanpour, S.S. Gill, A.N. Toosi. “Performance evalution metrics for cloud, fog, and edge computing: a review, taxonomy, benchmarks and standards for future research,” Internet of Things, vol 12. pp. 100273, 2020.
[24] F. Arute *et al.,* “Quantum supremacy using a programmable superconducting processor,” *Nature*, vol. 574**,** pp. 505–510, 2019.
[25] A.J. Ayer. *Language, Truth, and Logic*. Dover Publications, 1952, p. 45.
[26] E.-J. Wagenmakers, G. Dutilh, A Sarafoglou, “The creativity-verification cycle in psychological science: new methods to combat old idols,” *Perspectives on Psychological Science*, vol. 13, no. 4, pp. 418-427, 2018.
[27] https://irds.ieee.org
[28] https://www.src.org/about/decadal-plan/
[29] M.E. Oswald and S. Grosjean, “Confirmation Bias,” *Cognitive Illusions*. Ed. R.F. Pohl (Psychology Press, 2004).
[30] https://moore.org/about/founders-intent accessed 4/19/2021.
[31] Rebooting the IT Revolution (2015).
[32] A. de Vries, “The growing energy footprint of artificial intelligence,” *Joule* vol. 7, no. 10, pp. 2191-2194, 2023.
[33] https://www.iea.org/reports/electricity-2024 accessed 9/24/2024.
[34] R.F. Barrett *et al.*, “On the role of co-design in high performance computing,” *Transition of HPC Towards Exascale computing*. Ed. E.H. D’Hollander, J.J. Dongarra, I.T. Foster. (IOS Press, 2013).
[35] Z. Tokei *et al*., “Inflection points in interconnect research and trends for 2nm and beyond in order to solve the RC bottleneck.” *2020 IEEE International Electron Device Meeting*, 2020, pp. 32.2.1-32.2.4
[36] M. Bamal *et al.*, “Performance comparison of interconnect technology and architecture options for deep submicron technology nodes,” *2006 International Interconnect Technology Conference*, pp. 202-204, 2006.
[37] R. Kirchain and L. Kimerling, “A roadmap for nanophotonics,” *Nature* Photonics, vol. 1, pp. 303-305, 2007.
[38] J. Hecht, “The bandwidth bottleneck,” *Nature*, vol. 536, pp. 139-142, 2016.
[39] D. Efnusheva, A. Cholakoska, and A. Tentov, “A survey of different approaches for overcoming the processor-memory bottleneck,” *International Journal of Computer Science and Information Technology*, vol. 9, no. 2, pp. 151-163, 2017.
[40] P.A. Gargini, “How to successfully overcome inflection points, or long live Moore’s law,” *Computing in Science and Engineering*, vol. 19, no. 2, pp.51-62, 2017.
[41] S.K. Moore, “A Better Way to Measure Progress in Semiconductors,” *IEEE Spectrum*, vol. 10, pp. 24-28, 2020.
[42] T.M. Conte, I.T. Foster, W. Gropp, and M.D. Hill, “Advancing computing’s foundation of US industry and society.” https://cra.org/ccc/resources/ccc-led-whitepapers/#2020-quadrennial-papers
[43] S. Manipatruni *et al.*, “Scalable energy-efficient magnetoelectric spin–orbit logic,” *Nature*, vol. 565, pp. 35–42, 2019.
[44] G. Srinivasan, A. Sengupta, and K. Roy, “Magnetic Tunnel Junction Based Long-Term Short-Term Stochastic Synapse for a Spiking Neural Network with On-Chip STDP Learning,” *Scientific Reports* vol. 6, p. 29545, 2016.
[45] I. Chakraborty, A. Jaiswal, A.K. Saha, S.K. Gupta, and K. Roy, “Pathways to efficient neuromorphic computing with non-volatile memory technologies,” *Applied Physics Letters*, vol. 7, p. 021308, 2020.
[46] G. Finocchio, *et al.* “Roadmap for unconventional computing with nanotechnology.” *Nanofutures*, vol. 8. pp. 012001, 2004.
[47] M.M. Sabry Aly, *et al*. “Energy-efficient abundant-data computing: The N3XT 1,000x,” *Computer*, vol. 48. No. 12, pp. 24-33, 2015.
[48]https://eps.ieee.org/technology/heterogeneous-integration-roadmap/2024-edition.html
[49] Donglin Lu, *et al*., “Monolithic three-dimensional tier-by-tier integration via van der Waals lamination,” *Nature*, vol. 630, pp. 340-345, 2024.
[50] Yongkai Li, et al. “Miura-ori enabled stretchable ciruit boards,” *npj Flexible Electronics*, vol. 5, no. 3. pp. 1-9, 2021.
[51] C. Choi, *et al*., “Reconfigurable heterogeneous integration using stackable chips with embedded artificial intelligence,” *Nature Electronics*, vol. 5, pp. 386-393, 2022.
[52] D. Silver *et al.*, “Mastering the game of Go with deep neural networks and tree search,” *Nature*, vol. 529, pp. 484–489, 2016.
[53] C. Lee *et al*., “Human vs. Computer Go: Review and Prospect,” *IEEE Computational Intelligence Magazine*, vol. 11, no. 3, pp. 67-72, 2016.
[54] “Showdown,” *The Economist*. March 12, 2016.
[55] E. von Hippel. *Democratizing innovation*. MIT Press, 2005.
[56] M. Boroush, “US R&D Increased by $51 Billion to $606 Billion in 2018; estimate for 2019 indicates a further rise to $656 Billion,” *NSF National Center for Science and Engineering Statistics InfoBrief*, April 2021.
[57] J.V. Kennedy, “The sources and uses of U.S. science funding,” *The New Atlantis*, vol. 36, pp. 3-22, 2012.
[58] S. Bhatti *et al.*, “Spintronics based random access memory: a review,” *Materials Today*, vol. 20, no. 9, pp. 530-548, 2017.
[59] Y. Alexeev *et al.*, “Quantum computer systems for scientific discovery,” *PRX Quantum*, vol. 2, p. 017001, 2021.
[60] B. Godin, “The knowledge-based economy: conceptional framework or buzzword,” *Journal of Technology Transfer*, vol. 31, pp. 17-30, 2006.
[61] Kant’s quote: *Ignavum, fucos, pecus a praesepibus arcent*-Virgil, *Georgica*, IV. 168.
[Translation] They protect the hive from the drones, an idle bunch